\documentclass[sigconf]{acmart}

\AtBeginDocument{%
  \providecommand\BibTeX{{%
    \normalfont B\kern-0.5em{\scshape i\kern-0.25em b}\kern-0.8em\TeX}}}
    
\copyrightyear{2025} 
\acmYear{2025} 
\setcopyright{cc}
\setcctype{by}
\acmConference[CHI '25]{CHI Conference on Human Factors in Computing Systems}{April
26-May 1, 2025}{Yokohama, Japan}
\acmBooktitle{CHI Conference on Human Factors in Computing Systems (CHI '25), April
26-May 1, 2025, Yokohama, Japan}\acmDOI{10.1145/3706598.3713703}
\acmISBN{979-8-4007-1394-1/25/04}

\usepackage{natbib}
\usepackage{graphicx}
\usepackage{float}
\usepackage{colortbl}
\usepackage[utf8]{inputenc}

 \usepackage{xcolor}

 \definecolor{pedagogy}{HTML}{37B5BC} 
 \definecolor{action}{HTML}{4D9961} 
  \definecolor{action}{HTML}{4D9961}
  \definecolor{practice}{HTML}{998E25}
  \definecolor{engage}{HTML}{AF5454}
  \definecolor{explore}{HTML}{5C5CAA}
  \definecolor{valid}{HTML}{317FAF}

\makeatletter
\renewcommand{\@secnumfont}{\bfseries} 
\renewcommand{\section}{\@startsection{section}{1}{\z@}%
  {-.75\baselineskip}{.25\baselineskip}%
  {\Large\bfseries\MakeUppercase}} 
\makeatother

\begin{document}


\title[Disentangling the Power Dynamics in Participatory Data Physicalisation]
      {Disentangling the Power Dynamics in  Participatory Data Physicalisation}


\author{Silvia Cazacu}
\email{silvia.cazacu-bucica@kuleuven.be}
\orcid{0000-0002-7952-0919}
\affiliation{%
  \institution{KU Leuven}
  \city{Leuven}
  \country{Belgium}}

\author{Georgia Panagiotidou}
\email{georgia.panagiotidou@kcl.ac.uk}
\orcid{0000-0003-4408-6371}
\affiliation{%
  \institution{King's College London}
  \city{London}
  \country{United Kingdom}}

  \author{Thérèse Steenberghen}
\email{therese.steenberghen@kuleuven.be}
\orcid{0000-0002-7693-434X}
\affiliation{%
  \institution{KU Leuven}
  \city{Leuven}
  \country{Belgium}}

\author{Andrew Vande Moere}
\email{andrew.vandemoere@kuleuven.be}
\orcid{0000-0002-0085-4941}
\affiliation{%
  \institution{Research[x]Design - KU Leuven}
  \city{Leuven}
  \country{Belgium}}


\begin{abstract}
Participatory data physicalisation (PDP) is recognised for its potential to support data-driven decisions among stakeholders who collaboratively construct physical elements into commonly insightful visualisations. 
Like all participatory processes, PDP is however influenced by underlying power dynamics that might lead to issues regarding extractive participation, marginalisation, or exclusion, among others.
We first identified the decisions behind these power dynamics by developing an ontology that synthesises critical theoretical insights from both visualisation and participatory design research, which were then 
systematically applied unto a representative corpus of 23 PDP artefacts. 
By revealing how shared decisions are guided by different agendas, this paper presents three contributions: 
1) a cross-disciplinary ontology that facilitates the systematic analysis of existing and novel PDP artefacts and processes; which leads to
2) six PDP agendas that reflect the key power dynamics in current PDP practice, revealing the diversity of orientations towards stakeholder participation in PDP practice; and
3) a set of critical considerations that should guide how power dynamics can be balanced, such as by reflecting on how issues are represented, data is contextualised, participants express their meanings, and how participants can dissent with flexible artefact construction.
Consequently, this study advances a feminist research agenda by guiding researchers and practitioners in openly reflecting on and sharing responsibilities in data physicalisation and participatory data visualisation.
\end{abstract}



\begin{teaserfigure}
  \includegraphics[width=\textwidth]{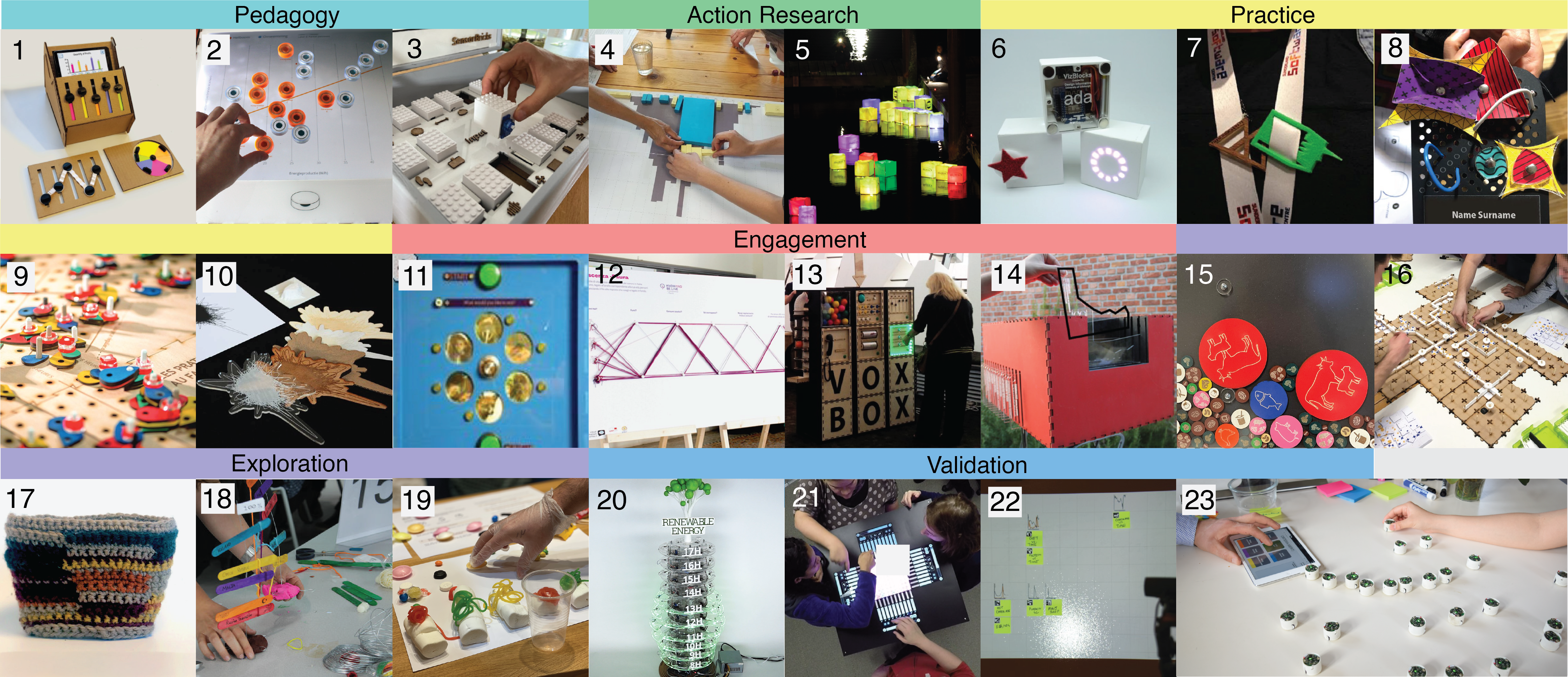}
  \caption{By analysing a corpus of 23 PDP artefacts, we synthesised six agendas that highlight different orientations towards using PDP as a participatory method: \textit{\textcolor{pedagogy}{Pedagogy}}, \textit{\textcolor{action}{Action Research}}, \textit{\textcolor{practice}{Practice}}, \textit{\textcolor{engage}{Engagement}}, \textit{\textcolor{explore}{Exploration}}, \textit{\textcolor{valid}{Validation.}} Image credits belong to the respective authors.}
  \Description{This teaser image showcases 23 PDP artefacts organised into six categories: Pedagogy, Action Research, Practice, Engagement, Exploration, and Validation. Pedagogy: Includes artefacts such as a physical toolkit with colourful components (1), a grid with coloured discs representing data points (2), and a LEGO-based model for data interaction (3). Action Research: Features hands manipulating various physical models—building blocks on a paper grid (4), glowing cubes floating on water at night (5), and interactive boxes with lights and tactile components (6). Practice: Displays interactive wearable devices, such as custom badge-making kits (7) and geometric models for organising data visually (8). Engagement: Visualises interactive installations where participants engage with physical representations of data, such as multi-coloured buttons and pins (9), acrylic shapes resembling leaves (10), and a coin-operated panel with visual data feedback (11-13). Exploration: Represents a diverse array of artefacts like a crocheted data visualisation (17), a map with small blocks placed on key points (18), and physical stacks that represent renewable energy usage (20). Validation: Highlights hands-on group interactions with physical tools and digital screens for validation of data points (21-23).}
  \label{fig:teaser}
\end{teaserfigure} 


\maketitle

\section{Introduction} \label{sec:introduction}
Data physicalisation involves the construction of 
\textit{ `a physical artefact whose geometry or material properties encode data'} \cite{jansen_opportunities_2015}.
Interacting with such a physical artefact enhances the cognitive understanding of data \cite{jansen_evaluating_2013}, as an individual can more easily relate the physical affordances and interaction mechanisms \cite{sauve_physecology_2022} into their personal experiences \cite{bressa_whats_2022}. By embodying data in a way that is both relatable \cite{zhao_embodiment_2008} and aesthetically pleasing \cite{hogan_towards_2017}, a data physicalisation can be experienced through multiple senses \cite{hogan2012does}, evoke enjoyment \cite{hogan2018data}, and even engage the body as a whole \cite{hornecker_design_2023}. For an individual, the process of constructing a data physicalisation can become a profound experience \cite{huron_constructing_2014}, as it allows them to explore and convey meanings that are otherwise challenging to measure or quantify \cite{van_koningsbruggen_metaphors_2024}. This process not only fosters creative data exploration \cite{bae_cultivating_2023}, but also empowers individuals to address personal needs \cite{bae_making_2022}, and express their concerns tangibly \cite{thudt_self-reflection_2018}. 
The natural affordances of data physicalisation enable individuals to integrate data in their daily life \cite{bhargava_data_2017}, empowering them to meaningfully participate in various complex data practices like data creation, data analysis, and data-driven decision-making \cite{neff2017critique}.
 
By bringing multiple individuals together to construct a shared artifact, data physicalisation also facilitates collaborative data practices.
Through this shared construction process, individuals align their personal meanings into a collective, data-driven narrative \cite{lockton_thinking_2020} enabling discussions about their individual inputs with one another \cite{sauve_edo_2023} and fostering negotiations on how their personal meanings contribute to the group narrative \cite{panagiotidou_supporting_2023}.
These collaborative aspects have allowed data physicalisation to address various civic issues by fostering participatory decision-making. For example, it has been used to combat environmental pollution \cite{perovich_chemicals_2021}, promote health awareness \cite{moretti_participatory_2020}, and investigate energy availability \cite{daniel_cairnform_2019}. Such applications often take place through participatory activities, such as mental mapping workshops \cite{panagiotidou_co-gnito_2022}, where participants collectively visualise data related to their surroundings; or prototyping sessions \cite{bae_cultivating_2023} where they build physical artefacts together. Several community engagement events \cite{gourlet_cairn_2017} and community action initiatives \cite{perovich_chemicals_2021} also leveraged data physicalisation as an inclusive data practice, enabling directly impacted stake-holders to engage with and make informed decisions about local challenges.
Among the diverse stakeholders who already integrated physicalisation into their data practices are local communities \cite{perovich_chemicals_2021}, schools \cite{veldhuis_coda_2020,bae_cultivating_2023}, non-profit organisations \cite{panagiotidou_supporting_2023}, and even local enterprises \cite{lechelt_vizblocks_2023}.

These collaborative practices have been defined as participatory data physicalisation (PDP) -\textit{`a physical visualisation that allows a co-located audience to physically participate in the creation of the visualisation by directly encoding their data while following predetermined rules'} \cite{sauve_edo_2023} or \textit{`a collaborative physical representation of data by multiple people'} \cite{participation2022}.

The participatory process of constructing a data physicalisation aligns with \textit{participatory design}, as it similarly brings together multiple stakeholders in making decisions \cite{halskov2015diversity} around developing a common solution for a fuzzy, yet meaningful, problem by way of creative \cite{cross2011design} and iterative prototyping \cite{Cross2006}. Participatory design (PD) is an established research discipline that aims to empower individuals to shape issues of their concern \cite{bodker_participatory_2018}, such as by developing tools and methods \cite{spinuzzi2005methodology} that foster creativity and collaboration \cite{sanders_probes_2014, sanders_co-creation_2008}. 
Research in PD has however shown that the involvement of stakeholders in decision-making is contingent on their agendas \cite{bodker_participation_2015} and the stakeholders' authoritative power to make decisions on behalf of others \cite{bodker_tying_2017}, while the designers that plan the participatory process exert power through their expertise \cite{bratteteig2012disentangling}. 

This presence of power dynamics carries significant implications, particularly when the stakeholder contributions are not reciprocated with a truly transformative outcome \cite{teli_tales_2020,huybrechts_visions_2020}. Moreover, these power dynamics can even lead to marginalisation and exclusion concerns \cite{beck2002p}, as participatory design activities are often disproportionately attended by individuals from dominant social categories \cite{bjorgvinsson2012agonistic}.
These risks can become exacerbated when participation occurs within the context of a data practice, as the production and use of data are influenced by decisions \cite{akbaba2024entanglements} made within specific social, infrastructural, and local contexts \cite{loukissas_open_2021}. These decisions have been shown to prioritise universalist, Western-dominated perspectives \cite{milan2019big} and anthropocentric views \cite{sheikh2023more, akama_expanding_2020}. Considering the complexities that arise from power dynamics in participation and the context of data, we argue that PDP must be examined as a participatory data practice.

Building on the understanding that participatory data practices are inevitably influenced by power dynamics, 
we define PDP as a \textit{participatory method where stakeholders collaboratively construct a physical representation that embodies shared meanings in data.}
This interpretation underscores how PDP is inherently shaped by power dynamics that emerge from the decisions made during its collaborative construction process, which are then embedded in the resulting data physicalisation artefact. We thus structure our analysis around the following research questions: \textit{What are the power dynamics between stakeholders and designers of PDP? How do these power dynamics influence PDP planning, deployment and construction?}

We believe that data feminism offers an appropriate lens to examine 
these research questions because it challenges the perceived z of data practices by revealing the social and technological contexts in which data is created and used \cite{dignazio_data_2020}.
We thus synthesised contributions from both participatory design and visualisation literature that, in our interpretation, align with data feminist values of transparency and empowerment by engaging with perspectives of situated and embodied knowledge.

Therefore, we operationalised these theoretical contributions by building a cross-disciplinary ontology that captures knowledge from both disciplines, which we structured along five fundamental dimensions: \texttt{WHO, WHAT, WHY, WHEN/WHERE} and \texttt{HOW}. 
We then applied this ontology through a codebook thematic analysis \cite{braun_conceptual_2022} 
to a representative corpus of 23 PDP artefacts that were identified via a systematic literature review. 
This analysis resulted in a series of themes, sub-themes and codes that revealed the context in which power dynamics operate when multiple individuals are invited to construct PDPs collaboratively.
By then reflecting upon these themes, sub-themes and codes, this paper presents three contributions: 
1) a cross-disciplinary ontology that facilitates the systematic analysis of existing and novel PDP artefacts and processes; which leads to
2) six PDP agendas that reflect the key power dynamics in current PDP practice, revealing the diversity of orientations towards stakeholder participation in PDP practice; and
3) a set of critical considerations that should guide how power dynamics can be balanced, such as by reflecting on how issues are represented, data is contextualised, participants express their meanings, and how participants can dissent with flexible artefact construction.
Accordingly, this study advances a feminist research agenda by informing researchers and practitioners how to more openly reflect on and share their responsibilities when using data physicalisation in particular, and data visualisation within a participatory context, in general.

\section{Background} \label{sec:background}

From a data feminist perspective, data physicalisation can be used as a tactic by educators who want to teach novices how to work with data in a playful and intuitive way \cite{d2017creative}, because by physicalizing it, they perceive data as a material that they can easily manipulate to capture the phenomenon they want to make claims about \cite{offenhuber2020we}. For example, by manipulating physical blocks whose size represents the amount of energy used by appliances, individuals who otherwise might have difficulty interpreting data, can now easily plan the daily energy consumption within their household \cite{panagiotidou_supporting_2023}.
As this example showed, the construction of physicalisations engages the entire human body in making sense of data, which builds on situated \cite{vdmoere2012designing} and embedded visualisations \cite{willett2016embedded} whose goal is to allow individuals to encounter data serendipitously in their everyday life. 
Furthermore, physicalisation expands the goal of personal visualisations \cite{huang2014personal} to broaden data analysis from professional settings to personal insights and communal interests, by supporting individuals to contribute data that is relevant to them and their communities.
Considering that many examples of physicalisations have already been documented for their ability to support individuals in data-based reflection \cite{bae_making_2022} and to encode new data \cite{inputvis24} in places~\cite{ bressa_whats_2022} and through interactions \cite{sauve_physecology_2022} that they find meaningful, we hypothesise that physicalisations are built on the premise that they offer a more transparent alternative for individuals to participate in meaningful data-driven activities by naturally exposing the decisions and points of view involved in their construction.

We use a feminist lens to unpack why decision-making processes in PDP should be made visible, recognising that feminist theory has long examined how decisions are inherently shaped by power and privilege. Several key feminist theories help to illuminate this point. Donna Haraway’s theory of situated knowledges critiqued the notion of scientific objectivity as a male-centric privilege and advocated for acknowledging context and perspective in decision-making as a way to reclaim power through knowledge \cite{haraway1988situated}. Kimberlé Crenshaw builds on this by demonstrating how intersecting identities expose the ways power operates through overlapping systems, such as race, class, and gender \cite{crenshaw2013mapping}. Patricia Hill Collins conceptualises these intersecting systems as a `matrix of domination,' showing how they collectively shape the lived experiences of marginalised groups \cite{collins1990black}. Extending this, Karen Barad’s theory of entanglement highlights how knowledge and power are co-produced through feedback loops that perpetuate systemic inequalities \cite{barad2007meeting}. Together, these perspectives underscore the necessity of making decisions visible, as transparency allows us to confront and potentially dismantle the power imbalances embedded in participatory practices.
Considering these perspectives, in the remainder of this section we focus on power inherent to data as a means of knowledge production, and the power exerted by those that are actively involved in its creation process.

\subsection{Power in data practices} \label{sec:power in data}

As art historian Johanna Drucker posited, power can be exerted through data because it is not `given' but rather `taken' \cite{drucker2014graphesis}, emphasizing that data does not simply exist in the natural world but is instead an artefact constructed through means of knowledge production. Expanding on this perspective, Akbaba, Klein, and Meyer argue that data is produced through the interplay between the phenomenon it describes and the context in which it is generated, shaped by measuring devices, their limitations, and the decisions guiding the measurement process \cite{akbaba2024entanglements}.
Building on this understanding, data practices have been critiqued for promoting universalist, Western-dominated ideologies \cite{milan2019big} and anthropocentric discourses \cite{sheikh2023more, akama_expanding_2020}. This can be explained by how, in data visualisations, decisions are not always transparently communicated \cite{dimara2021critical}, and uncertainty in the data is sometimes deliberately concealed \cite{liu2019understanding, liu2020paths}, reflecting systemic deficiencies that normalise the omission of uncertain data \cite{hullman2019authors}.
Among these issues, it is important to acknowledge that visualisations are often created under challenging conditions, defined by complex power asymmetries between design teams and stakeholders. Designers must navigate constraints that extend beyond the scope of the visualisation itself, such as mentoring students or securing funding \cite{akbaba_troubling}, all while working with collaborators whose roles or expertise may be unclear at the outset, but who hold the power to influence or delay projects \cite{sedlmair2012design}.
In this context, often the labour of analysing, curating or archiving data remains hidden or unattributed, giving the false impression that data visualisations are finished products rather than the result of ongoing work \cite{correll_ethical_2019}. As a consequence, visualisations can marginalise or obscure alternative viewpoints and flatten diverse life experiences, potentially reinforcing dominant narratives and excluding critical perspectives \cite{dignazio_data_2020}. This impact is particularly significant for marginalised groups when visualisations are improperly maintained or misused \cite{mcnutt2020surfacing}.

In response, participation in today's society depends on individuals who are able to engage critically in data practices~\cite{dencik2019exploring} because it relies on making decisions informed by large amounts of data, which is typically produced by means controlled by powerful groups \cite{dignazio_data_2020}. This criticality gives individuals the ability to challenge unfair practices by engaging in tactics and strategies that repurpose data for social good \cite{pangrazio2023critical}; by having the awareness that data is shaped by social, infrastructural and local aspects \cite{loukissas_open_2021}; 
and the understanding that data is often experienced as an intersectional and communal asset by minoritised groups \cite{johnson_data-literacy_2021}.
Responsible and safe data practices are also consensual data practices \cite{garcia_critical-refusal}, where participants engage in an ongoing process of both giving and withdrawing consent \cite{im2021yes, garcia_critical-refusal,blackwell2017classification, bardzell2010feminist, dimond2013hollaback}. This concept applies not only to individuals but also to scientists who choose not to collect new data or generate and share new knowledge, as well as to activists who participate in collective efforts to advance data justice \cite{garcia_critical-refusal}.

Given how the creation and use of data are intricately intertwined with the social context where it is understood, represented, and acted upon, we consider that PDP practice, as a component of the broader data visualisation practice, should be critically examined as a mechanism for enacting power.

\subsection{Power in participatory design} \label{power in PD}

The concern of transparency in participatory practices is addressed as a key contribution of participatory design (PD) research, as it systematically uncovers the decision-making processes among individuals and groups working together to solve real-world issues of their concern \cite{frauenberger_pursuit_2015}. Starting as an early Scandinavian contribution to Human-Computer Interaction \cite{bodker_participatory_2018} (for an overview of PD contributions to HCI see \cite{rogers_hci_2022}), the field of PD has been initially driven by a socialist political commitment to empower workers to share the design decisions about emerging technologies that they were using in their work \cite{halskov2015diversity, bodker_participation_2015}. Over time, PD has evolved into a research methodology influenced by the principles of Action Research \cite{foth2006participatory}, thus it draws on research methods such as interviews and participant observations to inform the iterative design of technologies that further elicit research results, which are then interpreted together with participants who are the users of these technologies \cite{spinuzzi2005methodology}. Because participants are not always able to imagine solutions to problems addressed by design, PD researchers have developed a variety of tools, techniques and participation methods \cite{sanders_probes_2014} to support individuals to express and communicate their ideas \cite{sanders_co-creation_2008}.

But to be able to respond meaningfully to issues that impact a diversity of groups, the PD process go to great lengths to not only approach participants, but to orchestrate long-term collaborations between researchers and a range of public~\cite{boone2023data}, academic \cite{peer2019workshops} and non-profit \cite{boone2023data} stakeholders. As such, PD approaches the design process not in a research lab, but within a network of stakeholders with intricate relationships \cite{gartner1996mapping} that affect the way issues are identified and addressed. Revealing how these relationships influence the power dynamics between stakeholders represents thus the fundamental ideal of PD to support reflexive design processes that question authority and give voice to marginalised groups~\cite{bardzell2018utopias}. 
Stakeholders exercise power in a design process through decision-making. It has been shown that decisions such as those for value alignment within a design process, decisions for implementing common visions or resolving conflicting ones, and decisions leading negotiations with the external world of a design process, are typically entangled with different manifestations of power, such as influence, ability, or expertise \cite{bratteteig2012disentangling}.
The entanglement of power and shared decisions represents the core of participation in PD \cite{bratteteig_unpacking_2016}, since the two components of power, as theorised by Hanna Pitkin \cite{pitkin1973wittgenstein} as \textit{`power to'} and \textit{`power over'}, affect how decisions are made and vice-versa. We build on this work to unpack how the decisions made by different stakeholders affect the construction process of PDP and the resulting artefact.

The \textit{`power to'} describes the ability and the agency of a person to (en)act decisions in order to reach the outcomes that they intend. This is the type of power that PD typically aims to support participants in gaining. For instance, when volunteers to a local food bank design visualisations of warehouse activities, they enact decisions to prioritise certain the shipment routes, thus gaining the ability to plan the process of food supply \cite{rossitto_efficiency_2021}.
\textit{`Power over'} refers to having an influence over decisions that others might make in the design process, thus it is an indirect type of power that typically stakeholders have in participation process. For example, participatory workshops where teachers design new educational technologies for schools, are typically influenced by political and practical decisions of the organisations in the educational system where those teachers work \cite{bodker_tying_2017}.
Another important influence over decisions represent agendas i.e. the priorities that motivate and legitimise stakeholders decisions \cite{gartner1996mapping} which determine the scope and approach towards the issues addressed \cite{frauenberger_pursuit_2015}.

These two types of power do not manifest in isolation from each other, since stakeholders and participants exert both types in their decision-making process. 
When participants enact their \textit{power to}, they develop new skills by actively engaging in the iterative and collaborative design process. Through iterative practice and dialogue with others, they learn to articulate their viewpoints on the issue they are addressing \cite{frauenberger_pursuit_2015}. These newly developed abilities to communicate effectively their perspectives, in turn, support their power to influence other stakeholders on the very issues that they care about \cite{frauenberger_scale_2018}.
Apart from participants and stakeholders, PD researchers also exercise power through their expertise of eliciting knowledge about the issues addressed from other stakeholders \cite{bratteteig2012disentangling} and their capacity of planning participation projects \cite{bodker_participation_2015}. Their expertise and leadership helps them build influence over who gets to participate, how the participation process is further deployed and the intended outcomes previously set by agendas \cite{bodker_participatory_2018}.
To all these power manifestations through active decision-making, the power of delaying or omitting decisions altogether adds a layer of bias or ambiguity to design ~\cite{bratteteig2012disentangling}. By addressing issues in a biased or controversial manner \cite{baibarac-duignan_controversing_2021}, or by allowing for criticality and ambiguity in the design process \cite{kaethler2017ambiguity}, both designers and stakeholders can steer a project towards their intended agenda by prioritising or delaying actions.

Building on biased, ambiguous or delayed decisions, intersectional power \cite{crenshaw2013mapping} adds new layers of complexity to power dynamics in participation. One of the issues that PD researchers should be aware is the intersection between participants' skill-sets and their identities. By changing from a context where one is a minority to a context where one is an expert in control, the researchers have the power to rob participants of the confidence in their abilities to participate~\cite{power_differentials}. This overview of power dynamics in PD practices demonstrates why PDP processes should be submitted to the same level of scrutiny as the outcome.

By examining the dynamics of power that emerge when individuals with different identities, values and abilities get together to construct data physicalisations, this paper answers the calls for more methodological diversity in visualisation practice \cite{losev2022embracing} by valuing its generative and exploratory attributes \cite{hinrichs2018defense}, by establishing it as a rigorous form of knowledge production~\cite{meyer_criteria_2019}, and by prioritising transferability over reproducibility of visualisation results \cite{sedlmair2012design}. As such, our methodology builds on prior theoretical efforts from two distinct disciplines of participatory design and visualisation to establish a set of themes that help us analyse existing PDP practices considering the aspects of (physical) visualisation and participation, both equally significant and mutually informing.

\begin{figure*} 
    \centering
    \includegraphics[width=\textwidth]{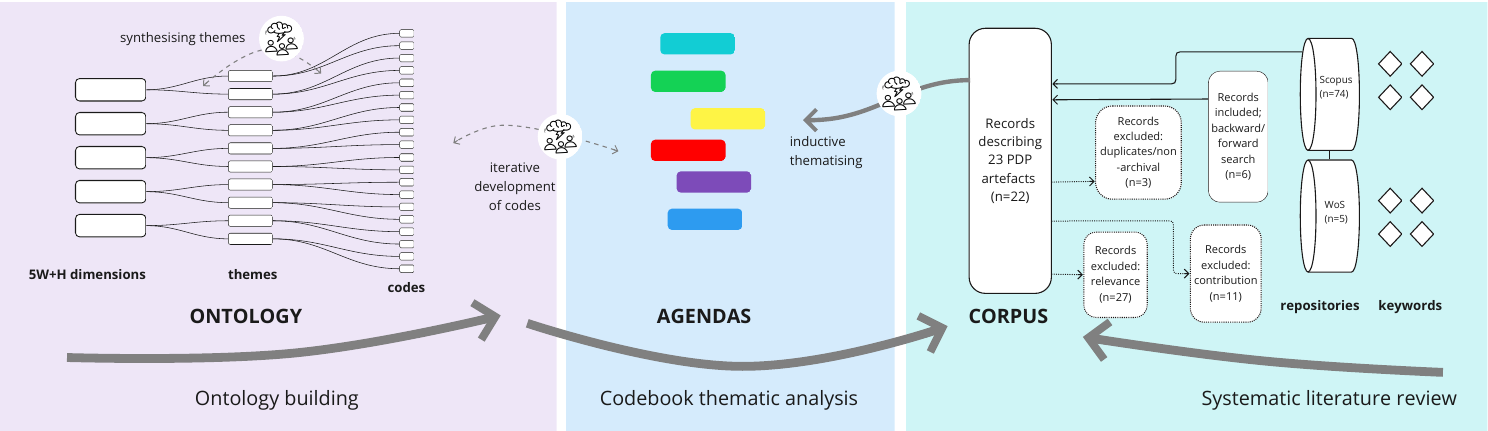}
    \caption{To highlight how power dynamics influence PDP practice, we synthesised six PDP agendas. From left to right: in \textit{Ontology building}, we iteratively synthesised 5W+H dimensions with corresponding themes, sub-themes and codes. From right to left: we developed a representative corpus by conducting a \textit{Systematic literature review}. In the middle, we performed a \textit{Codebook Thematic Analysis} of the Corpus by applying the Ontology.}
    \Description{The methodology image depicts the three distinct phases of the research process with separate coloured boxes. The boxes are connected by arrows to indicate the flow from one phase to the next, starting from right to left, then from left to right and finally arriving at the middle for the third phase. The flowchart uses icons to depict the output of each phase.}
    \label{fig:methodology}
\end{figure*}

\section{Methodology}

As shown in Figure \ref{fig:methodology}, our methodology combines inductive and deductive reasoning across three phases. In the first phase (Section \ref{sec:ontology_building}), we drew on the authors' collective expertise in visualisation and participatory design to select and analyse theoretical contributions that foreground the context where decisions are taken in PDP, in line with feminist calls for transparency \cite{dignazio_data_2020}.
However, this subjective literature selection introduced selective bias into the analysis.
To address this, the second phase (Section \ref{sec:systematic}) employed a systematic, keyword-based approach to compile a diverse corpus of exemplary PDP artefacts from academic literature. This ensured inclusivity across research communities reporting on designs that foster shared meaning-making through the physicalisation of data. Building on the first two phases, we conducted a thematic analysis of the corpus using the literature from the first phase as a codebook, as detailed in Section \ref{sec:codebook_ta}.

\subsection{Ontology building} \label{sec:ontology_building}

We developed the ontology presented in Table \ref{results} and detailed in Appendix \ref{sec:ontology} by using an approach from Knowledge Management \cite{pinto_ontologies_2004}, which conceptualises an ontology as an approach to systematically and rigorously define domain-specific knowledge for fostering shared understanding across disciplines. Our ontology applies this approach by organising knowledge into a tree-like structure of dimensions, themes, sub-themes, and codes.
We followed a top-down strategy: first establishing dimensions, then identifying seed publications relevant to each dimension, and branching out themes and sub-themes based on definitions from these publications. Codes were created later through an iterative process, applying the ontology to the PDP artefact corpus and refining it through discussions on how it supports critical analysis of the selected PDPs.

\subsubsection{5W+H dimensions.}
We structured the ontology using the `5W+H' dimensions \cite{whetten_what_1989} of knowledge building because each dimension answers one of the five fundamental questions \texttt{WHO? WHAT? WHY? WHEN/WHERE?} and \texttt{HOW?} allowing us to build a wide understanding of the context where PDPs are planned, deployed and constructed. 
The \texttt{WHAT} set the boundaries of what is designed and constructed. 
The \texttt{HOW} question reveals the dynamic aspects involved in PDP as both data physicalisation and participation method.
By answering the \texttt{WHY} question, we unpacked the motivations behind decisions made by stakeholders, as revealed by the \texttt{WHO} question, in designing and constructing PDP. Finally, the \texttt{WHEN/WHERE} question positions PDP in relation to the activities, communities and interests of those stakeholders.

\subsubsection{Visualisation themes.}
Using subjective and objective criteria, we framed each of the five dimensions of the ontology in literature that provided answers to the five corresponding questions. 
By discussing with the expert co-authors with expertise in design and criticality, 
we subjectively identified six canon publications \cite{bae_making_2022, sauve_physecology_2022, hornecker_design_2023, bressa_whats_2022, kerzner_framework_2019, knoll_extending_2020} based on the co-authors collective knowledge and academic interests. We used these as seed publications because they aligned with the feminist research objectives of this study
by providing relevant and reflective contributions.

These publications examine data physicalisation in terms of an individual’s agency in knowledge creation \cite{bae_making_2022} \texttt{(WHAT)} and the implicit and experiential aspects that support meaning-making \cite{hornecker_design_2023} \texttt{(HOW)}. The discussions about public, semi-public and private situations \cite{sauve_physecology_2022} as well as the everyday lives and communities \cite{bressa_whats_2022} contribute a nuanced understanding of human-data interaction \texttt{(WHERE/WHEN)}. In collaborative contexts, these publications emphasise the need for reflection on facilitator roles \cite{kerzner_framework_2019}, and for trust in effective collaboration \cite{knoll_extending_2020} \texttt{(WHO, WHY)}.
Starting from these seed publications, we performed a backward and forward search which led us to identify five additional ones \cite{hullman2011visualisation, huron_constructing_2014, hogan_towards_2017, schoffelen2015visualising, sedlmair2012design} that covered relevant topics previously missing from our analysis: methods that add credibility to a dataset \cite{hullman2011visualisation}; construction techniques for encoding meaning \cite{huron_constructing_2014}; physicalisation artefact affordances \cite{hogan_towards_2017}; social and organisational values embedded in physical visualisations \cite{schoffelen2015visualising}; and misalignments between participants that influence visualisation outcomes \cite{sedlmair2012design}. We created a Miro board to visualise these literature contributions as we selected and grouped them.

To mitigate the subjective inclusion criteria in our search for relevant publications, we introduced objective criteria to only include high quality studies that were published in high impact venues, such as ACM CHI, ACM TOCHI and IEEE VIS, and that were systematically grounded in literature (such as \cite{bae_making_2022, hornecker_design_2023, bressa_whats_2022,sauve_physecology_2022} or in tens of use cases (such as \cite{kerzner_framework_2019, knoll_extending_2020}. Because they rigorously synthesised large amounts of information, these papers provided a diverse set of perspectives grounded in the examples they analysed.

\subsubsection{Participatory design themes.}
Using the same combination of subjective and objective selection criteria as with the visualisation themes, we identified five seed publications that, according to the expertise of the co-authors, provided relevant contributions on the following topics related to the political context of participatory decision-making:
conflicting stakeholder values \cite{frauenberger_pursuit_2015} \texttt{(WHO)}; geo-political constraints and implications \cite{huybrechts_visions_2020} \texttt{(WHERE/WHEN)}; participation impact \cite{hansen_how_2019} \texttt{(WHY)}; societal issues \cite{dombrowski_social_2016} \texttt{(WHAT)}; and PD processes \cite{bodker_tying_2017} and methods \cite{sanders_probes_2014} \texttt{(HOW)}.
Through backward and forward search we identified eight additional publications on the following topics: 
tensions due to unbalanced stakeholder agendas \cite{bodker_participation_2015}; 
designers' influence on PD processes ~\cite{bratteteig2012disentangling}; issues affecting everyday life \cite{latour2007reassembling}, evidenced with abstract data \cite{baibarac-duignan_controversing_2021}; PD outcomes \cite{falk_what_2021} and activities \cite{sanders_probes_2014}; community of purpose \cite{clarke_socio-materiality_2021}) and community of practice \cite{rossitto_efficiency_2021}.

Since the discipline of participatory design encompasses a wide range of methods designed for unique geo-political circumstances, groups, and goals, synthesising findings systematically does not necessarily contribute to the discipline in a meaningful way. Therefore, our objective selection instead targeted widely referenced publications that rigorously documented case studies \cite{frauenberger_pursuit_2015, huybrechts_visions_2020, sanders_probes_2014} and discussed their findings in connection to seminal literature contributions \cite{hansen_how_2019, dombrowski_social_2016}.

\subsubsection{Synthesising themes and codes.}
After grounding each dimension in at least one theme from visualisation and participatory design, we brainstormed over several months to merge, refine and expand the thematic representations and align interpretations among co-authors. 
During this process, we tested the ontology with 5 PDP artefacts from the corpus (see Section \ref{sec:systematic}), which led us harmonise theme definitions, allowing us to develop an encompassing definition for each of the five dimensions. 
We continued applying the harmonised ontology consisting of 11 seed publications and 13 additional publications to the remaining PDPs. By performing a thematic analysis (explained in Section \ref{sec:codebook_ta}), we developed sub-themes and codes that allowed us to describe the artefacts analysed. For instance, we divided the theme \textit{data collection} into the \textit{asynchronous} and \textit{synchronous} sub-themes, each with specific codes used to categorise different types of data collection activities, such as \textit{surveys}.

\subsection{Systematic literature review} \label{sec:systematic}
We built a corpus of 23 PDP artefacts by conducting a systematic literature review.

\subsubsection{Keyword-based search.}
According to our definition of PDP as a PD method, we started with a combinatory query of the terms \textit{participation} and \textit{data physicalisation} which only resulted in 10 records. This prompted us to widen the search with the related terms of \textit{collective}, \textit{collaborative}, \textit{workshop} and \textit{co-design} which described records that aligned with the intention of bringing together multiple individuals to construct a physical data representation. Although these new terms do not yield results that explicitly align with the empowerment values of PD processes, they did capture a diversity of physicalisation contributions. We also wanted to encapsulate broader interpretations of physicalisation thus included permutations like \textit{physicalisation}, \textit{representation}, \textit{materialisation}, and combinations of \textit{physical/tangible} and \textit{visualisation/materialisation} or \textit{representation}.

\subsubsection{Accessed repositories.}
Because we aimed to include contributions from diverse research communities that shared the same intentions of participatory physicalisation constructions, we conducted search queries on the \textit{Scopus} and \textit{Web of Science} repositories in July 2023.
We retrieved 74 \textit{Scopus} records and 5 additional records from \textit{Web of Science}, to which we added 4 records from backward and forward searching. We also added 2 more records which were published in the subsequent months after the original query and placed all records in an Excel worksheet to start the screening procedure.

\subsubsection{Exclusion criteria.}
After screening the title, abstract and keywords, 27 records were excluded because they did not engage with the intention to involve multiple individuals at any point before, during or after the process of physicalisation construction. Moreover, 19 records were excluded because their construction process did not result in an identifiable artefact (e.g. by presenting the results of a data physicalisation workshop), 11 because they did not contain mature and substantial research contributions (e.g. poster publications), 2 because they were non-archival publications and 1 because it was a duplicate record.
Despite the poster exclusion criteria, two short conference papers were kept since they reported on seemingly relevant innovations like the use of candy \cite{lallemand_candy_2022} and the collaboration between academia and creative industry \cite{lechelt_vizblocks_2023}. Multiple artefacts that were conjunctively discussed in a single record \cite{huron_lets_2017,brombacher_sensorbricks_2024} were considered singular, for the exception of one record that discussed two data physicalisation artefacts, one as part of a conference and one as part of a series of workshops \cite{nissen_data-things_2015}. As our research concentrated on mature academic contributions, our corpus intentionally comprises records detailing artefacts, given that academic papers generally document the associated construction processes. After applying all exclusion criteria, we did not identify records where the construction process has been undocumented. This phase resulted in a representative corpus of 23 PDP artefacts which we were able to analyse thematically in the following phase.

\subsection{Codebook thematic analysis} \label{sec:codebook_ta}
We analysed the corpus as constructed in Section \ref{sec:systematic} by applying the cross-disciplinary ontology developed according to Section \ref{sec:ontology_building} following a codebook protocol for thematic analysis \cite{braun_conceptual_2022}. By building a top-down analysis where themes are created a priori, this protocol supported a systematic analysis, while recognising that the positions of the researchers who performed the analysis influenced its result. The first author took the lead in coding the PDP artefacts, by applying one ontological dimension at a time onto an initial subset of 5 artefacts, and then continuing with new subsets of 5 until the entire corpus was covered. After each coded subset, the first author discussed the codes with the other co-authors to ensure that their interpretations were accurately captured and aligned. While the first subset of artefacts required several meetings during each the codes and sub-themes were iterated upon until we reached consensus, the remaining subsets generated only a few new codes. This approach allowed us to enriched the deductive protocol with the perspectives of a total of 4 researchers, including 2 experts. We present the results of this analysis in Section \ref{results section}, grouped according to the five ontological dimensions. 

Once the complexity of our analysis grew, the first author switched from Miro to an Excel spreadsheet that captured all artefacts analysed according to ontology structure presented in Table \ref{results}. This allowed us to zoom out from the deductive coding process to be able to discuss similarities between the PDP goals across dimensions, visible when several PDP artefacts used the same line of codes in Table \ref{results}.
We then thematised these artefacts collaboratively, based on similar aspects that define the decisions of stakeholders of PDP, as described by the themes and codes into common `agendas'.
When we were not sure whether an artefact belonged to a PDP agenda, we went back to the original publication that documented it to uncover which grouping was most fitting. As many PDP artefacts fit more than just one agenda, our categorisation is fluid rather than rigid so that it reveals the most prominent decisions influencing these PDPs, as described in Section \ref{agendas}. A capture of the Miro board and the spreadsheet are available at \url{https://doi.org/10.48804/CJM4L3}.

\section{Results} \label{results section}

In this section, we reveal how the relative distribution of decisions, shared by diverse stakeholders and designers influences the way PDP is used as a participatory method. We present these decisions grouped according to the five dimensions of the critical ontology - \texttt{WHO, WHAT, WHY, WHEN/WHERE} and \texttt{HOW} - and describe the 23 PDP artefacts from our corpus using the themes, sub-themes, and codes, as summarised in Table \ref{results}. A full description of the ontology terms used in this section can be consulted in Appendix \ref{sec:ontology}.

\begin{figure*}
    \centering
    \includegraphics[width=0.95\linewidth]{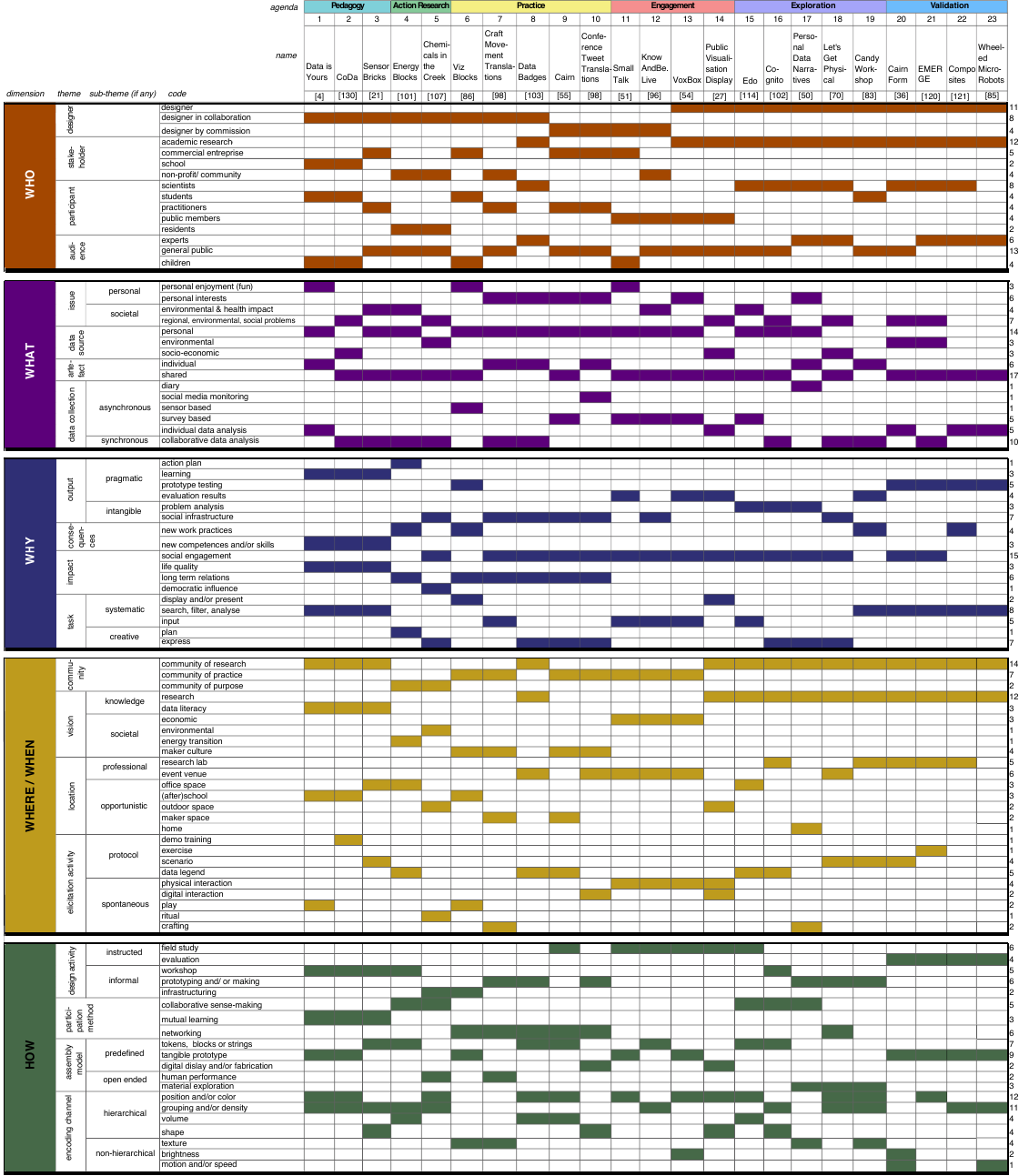}
    \captionsetup{type=table}
    \caption{ Our results categorise the decisions about the 23 PDPs in the corpus according to the 5 ontological dimensions WHO, WHAT, WHY, WHERE/WHEN and HOW and their respective themes, sub-themes and codes. Six agendas guide these decisions: \textit{\textcolor{pedagogy}{Pedagogy}, \textcolor{action}{Action Research}, \textcolor{practice}{Practice}, \textcolor{engage}{Engagement}, \textcolor{explore}{Exploration}} and \textit{\textcolor{valid}{Validation}}.}
    \Description{}
    \label{results}
\end{figure*}

\subsection{Who?} \label{results-WHO}
Our corpus indicates that decisions in PDP are generally distributed between two groups: designers who create PDPs independently to innovate within their academic research field (12, N=23), and designers who collaborate with local stakeholders, using physicalisation to address community-specific issues (11/23).

\paragraph{Designers as academic researchers.} 
In half of our corpus (12/23), designers created PDPs as part of their academic work in a research lab, dedicated to advancing the fields of visualisation and Human-Computer Interaction (HCI). In the visualisation contributions designers focused on exploring new techniques to physically represent data. For example, the \textit{Candy Workshop} \cite{lallemand_candy_2022} examines the sensory properties of candy to encode productivity data, while \textit{Edo} \cite{sauve_edo_2023} analyses how multiple individuals represent their eating habits in a shared physical artefact. In the HCI oriented PDPs, designers tested various interactive prototypes, including \textit{EMERGE} \cite{sturdee_exploring_2023}, a dynamic and actuated bar chart interface for collaborative data exploration, and \textit{CairnForm} \cite{daniel_cairnform_2019}, a shape-changing ring chart that represents data through motion.
Because this group of designers create novel PDP approaches in order to evaluate them, their physical data representations are addressed to an audience of experts (6/23) from within their immediate circle, knowledgeable about visualisation or interaction design, but not necessarily knowledgeable about the data being physicalised. As such, designers tend to favour a targeted involvement of scientists (8/23) to participate in their evaluation studies. One exception among these PDPs is \textit{Data Badges} \cite{panagiotidou_data_2020}, a PDP designed by academic researchers together with the organisers of a conference for visualisation experts, invited to physicalise data about their academic interests. 

\paragraph{Local stakeholders in collaboration with designers.} Ranging from commercial enterprises (5/23),  non-profit organisations or communities (4/23) and schools (2/23), local stakeholders enlist the expertise of physicalisation designers to communicate issues relevant for their communities.
Some of these stakeholders (4/23), mostly commercial enterprises (3/23), commission designers to create PDPs according to specific design briefs, such as to evaluate the experience of Fablab users, as seen in \textit{Cairn} \cite{gourlet_cairn_2017}. As a result, the local stakeholders are not involved in the design decisions, giving away their ability to receive bottom-up input from their own members, comprising of practitioners (2/4) and public members (2/4). 
Still, the majority of local stakeholders (8/23) act as subject matter experts who design PDPs in collaboration with expert designers. These stakeholders invite a variety of participants to construct PDPs, including students (3/8), practitioners (2/8) and residents (2/8).
In order to best communicate with their audience consisting of members of the general public (3/8), children (3/8), and experts (1/8), stakeholders share decisions together with designers. One example of collaboration is \textit{Chemical in the Creek} \cite{perovich_chemicals_2021}, where local NGOs team up with designers, academic researchers and students to raise awareness about the problem of water pollution through a community ritual that materialises open data about pollution incidents.  

\subsection{What?} \label{results-WHAT}
Our corpus suggests that the decisions regarding whether personal or societal issues are being physicalised influence how participants contribute their own data to the PDP artefact. 

\paragraph{Personal data about personal issues.} More than half of the PDPs in the corpus invite participants to encode personal data (14/23). Their data typically relates to issues of personal interest (6/14) such as DIY hobbies (\textit{Craft Movement Translations,  Conference Tweet Translations} \cite{nissen_data-things_2015} and \textit{Cairn} \cite{gourlet_cairn_2017}), academic research interests (\textit{Data Badges} \cite{panagiotidou_data_2020}), and daily commute to work (\textit{Personal Data Narratives} \cite{friske_entangling_2020}). Some personal data in PDPs relate to issues of personal enjoyment (3/14), encoding a fun activity such as a game in \textit{Data is Yours} \cite{bae_cultivating_2023}, or a children theatre play in \textit{Small Talk }\cite{gallacher_smalltalk_2016}. 
The personal data encoded is collected through a variety of synchronous (5/14) and asynchronous methods (9/14). 
Among the asynchronous data collection we notice surveys (5/14), social media monitoring in \textit{Conference Tweet Translations} \cite{nissen_data-things_2015}, sensors in \textit{VizBlocks} \cite{lechelt_vizblocks_2023} and diary study in\textit{ Personal Data Narratives} \cite{friske_entangling_2020}. The synchronous personal data collection always involves collaborative data analysis, where for example in \textit{Craft Movement Translations} \cite{nissen_data-things_2015} participants discuss their crochet patterns encoded into unique shapes by a movement sensor, and later laser-cut into physical artefacts.

\paragraph{Societal issues from different types of data.} Societal issues (11/23) were contributed from a mix of environmental data (3/11) and socio-economic data (3/11), but also personal data (5/11). These issues described environmental and health impact (4/23) including environmental impact of food choices in \textit{Edo} \cite{sauve_edo_2023}; and regional, environmental, social problems including water pollution.
All participants that are able to discuss societal issues (11/23) in first person, do so by constructing shared artefacts (11/11), most of which created synchronously (7/11).
For example, statistical environmental data was physicalised to represent in a participative way
energy forecasts by \textit{CairnForm} \cite{daniel_cairnform_2019}, water pollution by \textit{Chemicals in the Creek} \cite{perovich_chemicals_2021}, and average rainfall by \textit{EMERGE} \cite{sturdee_exploring_2023}. Because the construction occurred in a synchronous way, the underlying issues could be discussed by participants during collaborative data analysis.
The same approach was used when personal data about environmental and health issues in \textit{Energy Blocks} \cite{panagiotidou_supporting_2023} and \textit{SensorBricks} \cite{brombacher_sensorbricks_2024} was encoded by participants into shared artefacts by collaboratively comparing with others their personal behaviour.

\subsection{Why?} \label{results-WHY}

While the PDP construction is used to foster impact, as foregrounded in Section \ref{sec:ontology} as the long-term outcomes of the PD process on participants, stakeholders and society at large, through social engagement (15/23), the PDP artefact is used as a pragmatic output (13/23).
\paragraph{The PDP construction creates social engagement. }
The majority of our corpus aims to support social engagement as a way to create impact (15/23), engaging practitioners to exchange ideas about their shared DIY hobbies like \textit{Cairn} \cite{gourlet_cairn_2017} and \textit{Craft Movement Translations} \cite{nissen_data-things_2015} or to meet other academic researchers from the visualisation community by constructing \textit{Data Badges} \cite{panagiotidou_data_2020}.
Many of these socially engaging PDPs rely on intangible outputs (10/15) such as creating the premises of a new social infrastructure (7/10) by bringing together, often in the same space, individuals who share similar interests about future technologies, as in \textit{Conference Tweet Translations} \cite{nissen_data-things_2015} or who represent a local activism groups, like in \textit{Chemicals in the Creek} \cite{perovich_chemicals_2021}. Another way of creating intangible outputs is by supporting problem analysis (3/10) among participants who join to reflect on issues such as shared patterns of behaviour, as seen in \textit{Personal Data Narratives} \cite{friske_entangling_2020}. 
All these intangible outputs are meant not to create a physical artefact as an end in itself, but to use the PDP construction as an opportunity for individuals to join other likeminded ones in a collaborative reflection. This explains why these construction processes typically rely on creative physicalisation tasks such as asking participants to express (7/10) their experiences and emotions, as seen in \textit{Co-gnito }\cite{panagiotidou_co-gnito_2022} or in \textit{Let's Get Physical} \cite{huron_lets_2017}.

\paragraph{The PDP artefact as a pragmatic output.}
Most PDP artefacts used as a pragmatic output (13/23) for evaluation results (4/13) or for prototype testing (5/13) do not aim to achieve impact. A few exceptions were used with a clear intention to impact the life quality (3/13) or the long term relations of participants (2/13). These artefacts represent pragmatic outputs designed for learning, like \textit{Data is Yours} \cite{bae_cultivating_2023}, \textit{Coda} \cite{veldhuis_coda_2020}, and \textit{Sensor Bricks} \cite{brombacher_sensorbricks_2024} all designed to support participants to build new competences and skills; or \textit{Energy Blocks} \cite{panagiotidou_supporting_2023} designed to help participants devise an action plan about sharing energy consumption. 
The physicalisation tasks that participants are expected to fulfil during the construction of pragmatic artefacts are typically systematic (12/13), relying on activities such as search, filter and analyse with the exception of planning, a creative task fulfilled by the participants encoding their energy sharing habits (\textit{Energy Blocks} \cite{panagiotidou_supporting_2023}).

\subsection{When and where?} \label{results-WHEN}

Our corpus indicates that communities of research, practice, and purpose decide to implement their visions by constructing PDPs in locations and through elicitation activities that enhance the relevance of the issues being addressed.

\paragraph{Spontaneous versus protocol-based activities to implement visions.} PDPs incorporate spontaneous elicitation activities to integrate seamlessly data construction into participants' lives, as a way to implement societal visions (9/23) related to environmental impact (\textit{Chemicals in the Creek} \cite{perovich_chemicals_2021}), energy transition (\textit{Energy Blocks} \cite{panagiotidou_supporting_2023}), the maker culture (\textit{VizBlocks} \cite{lechelt_vizblocks_2023}, \textit{Craft Movement Translations} \cite{nissen_data-things_2015}, \textit{Conference Tweet Translations} \cite{nissen_data-things_2015}, \textit{Cairn} \cite{ gourlet_cairn_2017}). These PDPs elicit data from participants by engaging them in familiar activities including crocheting (\textit{Craft Movement Translations} \cite{nissen_data-things_2015}), observing a ritual (\textit{Chemicals in the Creek} \cite{perovich_chemicals_2021}), while enjoying a hobby (\textit{Cairn } \cite{gourlet_cairn_2017}).
Knowledge visions (14/23) advancing research are typically being carefully carried out by establishing thorough research protocols (11/14) involving demo training for \textit{CoDa} \cite{veldhuis_coda_2020}, exercises for \textit{EMERGE} \cite{sturdee_exploring_2023}, scenarios for \textit{Sensor Bricks} \cite{brombacher_sensorbricks_2024}, the \textit{Candy Workshop} \cite{ lallemand_candy_2022} and \textit{CairnForm } \cite{daniel_cairnform_2019}. This way, the researchers are able to control the data construction process up close. As an outlier, when aiming to support data literacy among participants, both academic research protocols such as a demo training for \textit{CoDa} \cite{veldhuis_coda_2020} and spontaneous play activities (\textit{Data is Yours} \cite{bae_cultivating_2023}) are prefered.

\paragraph{Communities favour specific locations.}
Communities of research implement their visions by inviting participants to the research labs, communities of practice organise public events, while communities of purpose go out into the field.
Members of communities of research (14/23) pursue their research vision by inviting participants to construct PDPs in their labs (5/14), as seen in the \textit{Candy Workshop} \cite{lallemand_candy_2022}, \textit{Co-gnito} \cite{panagiotidou_co-gnito_2022}, \textit{Cairnform }\cite{daniel_cairnform_2019}, \textit{EMERGE } \cite{sturdee_exploring_2023}, \textit{Composites }\cite{subramonyam_composites_2022} or at another professional location such as an event venue for academic researchers, as seen in the \textit{Let's Get Physical} workshop \cite{huron_lets_2017}. Still, they also construct PDPs in opportunistic places that are accessible to the target audience, such as \textit{Public visualisation Displays} set up in a city centre \cite{claes2015role}, or \textit{Edo} \cite{sauve_edo_2023} placed in the lunch area of a research office space. \textit{Personal data Narratives} \cite{friske_entangling_2020} are constructed at the participating researchers' homes.
In contrast to research, communities of practice (8/23) involve practitioners as participants, as seen in \textit{Data Badges} \cite{panagiotidou_data_2020}, \textit{Craft Movement Translations} \cite{nissen_data-things_2015} and \textit{Cairn} \cite{gourlet_cairn_2017}; and as organising stakeholders, as seen in K\textit{nowAndBe.Live }\cite{moretti_participatory_2020} and \textit{SmallTalk} \cite{gallacher_smalltalk_2016}. These communities set up PDP construction at public events (\textit{Data Badges} \cite{panagiotidou_data_2020}, \textit{VoxBox} \cite{golsteijn_voxbox_2015}, \textit{Conference Tweet Translations} \cite{nissen_data-things_2015}).
The least represented in our corpus, communities of purpose (2/23) tend to be organised around shared societal visions, such as water pollution in \textit{Chemicals in the Creek} \cite{perovich_chemicals_2021} or sustainable energy production in \textit{Energy Blocks} \cite{panagiotidou_supporting_2023}. 
They take a split approach to either construct PDPs in a public space (Chemicals in the Creek \cite{perovich_chemicals_2021}) or in an office space (Energy Blocks \cite{panagiotidou_supporting_2023}).

\subsection{How?} \label{results-HOW}

Our corpus shows that the decision to guide PDP construction through instructed design activities is being operationalised by a preference for predefined assemblies relying on encoding channels perceived as hierarchical. In contrast, informal activities rely on decisions to employ open-ended assemblies.

\paragraph{Improvised participation in instructed design activities.} Many PDPs improvised the way they involved participants into the construction process (10/23), rather than rely on a known participation method (8/10). Yet, they followed a predefined assembly model (7/8) to guide the participants during PDP construction.
For example, participants in the field studies of \textit{KnowAndBe.Live} \cite{moretti_participatory_2020} and \textit{VoxBox } \cite{ golsteijn_voxbox_2015} were not engaged through a specific participation method. Still, the construction of these PDPs was carefully steered through predefined assembly with strings and pins (\textit{KnowAndBe.Live} \cite{moretti_participatory_2020}) or the interaction with spinners and sliders of a tangible prototype (\textit{VoxBox} \cite{golsteijn_voxbox_2015}). These predefined assembly models are supported by hierarchical encoding channels where position and colour bring structure to the participants' visual mapping process, guiding them in the construction process.

\paragraph{Informal design activities to prioritise participation.} More than half of the artefacts in the corpus (13/23) were constructed by combining established participation methods, yet they still consolidated the construction with additional predefined assembly models (7/11). 
For example, all 5 workshops analysed (\textit{Data is Yours} \cite{bae_cultivating_2023}, \textit{CoDa }\cite{veldhuis_coda_2020}, \textit{SensorBricks} \cite{brombacher_sensorbricks_2024}, \textit{Energy Blocks }\cite{panagiotidou_supporting_2023}, \textit{Co-gnito} \cite{panagiotidou_co-gnito_2022}) relied on predefined assembly to provide structure to different participation methods. \textit{Data is Yours} \cite{bae_cultivating_2023}, \textit{CoDa }\cite{veldhuis_coda_2020} and \textit{SensorBricks} \cite{brombacher_sensorbricks_2024} supported mutual learning about data among students by relying on grouping blocks (\textit{Sensor Bricks} \cite{brombacher_sensorbricks_2024}) or interacting with tangible prototypes with digital display and tokens (\textit{Data is Yours} \cite{bae_cultivating_2023}, \textit{CoDa} \cite{veldhuis_coda_2020}). For \textit{Energy Blocks} \cite{panagiotidou_supporting_2023} and \textit{Co-gnito} \cite{panagiotidou_co-gnito_2022}, designers used blocks of different volumes and specific  configurations to facilitate collaborative sense-making among participants. All these methods used hierarchical encoding channels which inherently provided order to all the elements physicalised.
Although less prominent in our corpus, open-ended assembly (5/23) was also used for artefacts resulted from informal activities (5/5) most of them following a participation method (4/5). For instance, the human performance of hand-movement while crocheting was captured by motion sensors and later digitally fabricated into \textit{Craft Movement Translations} \cite{nissen_data-things_2015}. These individual artefacts supported participants to engage in networking with others about their shared craft.

\subsection{Agendas} \label{agendas}

\begin{figure}
    \centering
    \includegraphics[width=1\linewidth]{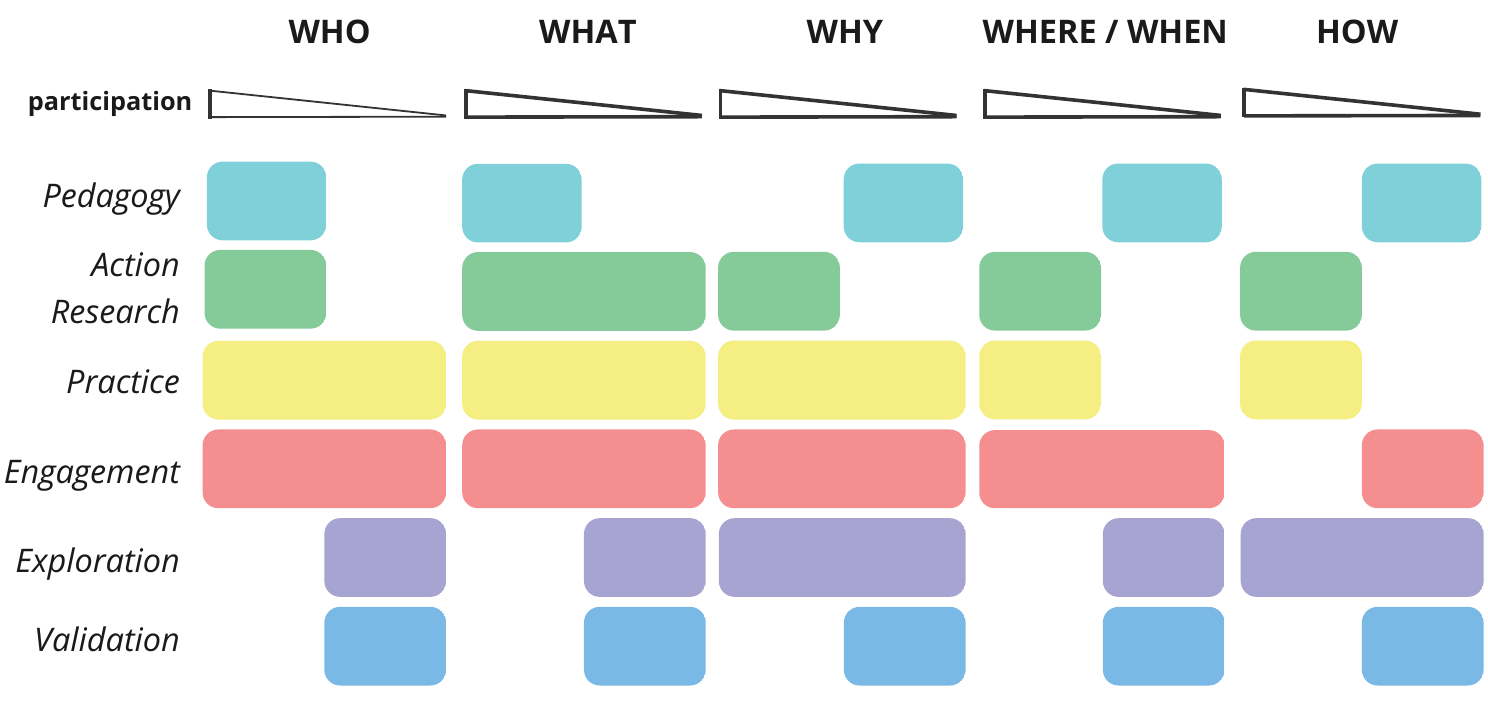}
    \caption{The six PDP agendas highlight different orientations towards participation. \textcolor{action}{\textit{Action Research}}, \textcolor{practice}{\textit{Practice}} and \textcolor{engage}{\textit{Engagement}} share a motivation to open the PDP planning and construction to the \textbf{participation} of stakeholders.
    \textcolor{pedagogy}{\textit{Pedagogy}}, 
    \textcolor{explore}{\textit{Exploration}} and \textcolor{valid}{\textit{Validation}} tend to prioritise innovative ways to embody meanings into data.}
    
    \Description{A figure illustrating six PDP agendas - `Pedagogy', `Action Research,' `Practice,' `Engagement,' `Exploration,' and `Validation' - each represented by a rectangle of a different colour, stacked vertically and aligned. The six agendas are repeated 5 times, each time representing one 5W+H dimension, positioned in a row.}
    \label{fig:agendas}
\end{figure}

Based on common aspects that define the decisions of stakeholders of PDP, as shown in Figure \ref{results}, we created six agendas (as previously defined in Section \ref{power in PD}) that guide the use of PDP as a participatory method: \textcolor{pedagogy}{\textit{Pedagogy}}, \textcolor{action}{\textit{Action Research}}, \textcolor{practice}{\textit{Practice}}, \textcolor{engage}{\textit{Engagement}}, \textcolor{explore}{\textit{Exploration}} and \textcolor{valid}{\textit{Validation}}. Many PDP artefacts analysed seem to encompass overlapping agendas, for example \textit{Data Badges} \cite{panagiotidou_data_2020} which capture a \textcolor{practice}{\textit{Practice}} agenda dedicated to engaging practitioners about shared interests and an \textcolor{explore}{\textit{Exploration}} agenda to explore how to create wearable PDPs. We briefly describe these six agendas, referencing the combined 5W+H ontological dimensions on which they are based and which determine the similar characteristics they share, as represented in Figure \ref{fig:agendas}.

\subsubsection{\textbf{Pedagogy}}

The PDPs whose main agenda is to support participants in building data literacy skills were grouped under \textcolor{pedagogy}{\textit{Pedagogy}}.
The intended output of their construction process is to help students and children learn basic data visualisation competences and skills so as to improve their data practices and in the long run, to positively impact their life quality in the digital age \texttt{(WHY)}. These PDPs are designed by academic researchers in collaboration with schools and commercial enterprises \texttt{(WHO)}. Therefore, they promote a knowledge vision related to data literacy, typically lead by communities of research \texttt{(WHERE/WHEN)}. The typically shared artefacts are built during individual or collaborative data analysis \texttt{(WHAT)}, thus they are used in workshop activities where they build on mutual learning as a participation method \texttt{(HOW)}.

\subsubsection{\textbf{Action research}}

The underlining agenda of \textcolor{action}{\textit{Action Research}} PDPs is to drive action and help communities address issues of their concern.
These PDPs are created by designers through long-term collaborations with non-profits \texttt{(WHO)} belonging to communities of purpose that pursue environmental \textit{Chemicals in the Creek} \cite{perovich_chemicals_2021} and energy transition \textit{Energy Blocks} \cite{panagiotidou_supporting_2023} visions \texttt{(WHERE/WHEN)}. Therefore, these communities aim to address societal issues of solar energy sharing and water pollution by collaboratively building shared artefacts in a synchronous process \texttt{(WHAT)}. The resulted PDP artefacts are typically deployed in opportunistic locations \texttt{(WHERE/WHEN)}, engaging participants using a participation method of collaborative sense-making \texttt{(HOW)}. Members of their communities are thus able to forge long term relations and even gain democratic influence \texttt{(WHY)}.

\subsubsection{\textbf{Practice}}

The main agenda of \textcolor{practice}{\textit{Practice}} PDPs is to bring practitioners together to engage with each other about their shared personal or professional interests related to data visualisation, digital technology, or DIY making. These PDPs are designed by communities of practice \texttt{(WHERE/WHEN)}, either through collaborative efforts between designers and local stakeholders or through projects commissioned by commercial enterprises \texttt{(WHO)}.
These PDPs prompt participants to contribute personal data to construct primarily shared artefacts \texttt{(WHAT)}, with the main objective of creating a social infrastructure that fosters long-term relationships among practitioners \texttt{(WHY)}. This is why they typically draw on participation methods centred on networking, while structuring the assembly process around predefined and hierarchical encoding \texttt{(HOW)}.

\subsubsection{\textbf{Engagement}}

The primary agenda of \textcolor{engage}{\textit{Engagement}} PDPs is to engage the public in issues that are personally and societally relevant including cancer awareness and local population statistics.
Lead by communities of research \texttt{(WHERE/WHEN)}, these PDPs are designed by academic researchers and sometimes commissioned by other stakeholders \texttt{(WHO)} to engage members of the public to create shared meanings by filling in surveys \texttt{(WHAT)} typically through the spontaneous interaction with a prototype placed at an event venue. Therefore the PDP construction relies on predefined assembly models and a combination of hierarchical and non-hierarchical encoding channels \texttt{(HOW)}.

\subsubsection{\textbf{Exploration}}

The PDPs whose main agenda is to explore new physicalisation techniques were grouped under \textcolor{explore}{\textit{Exploration}}.
Designed by academic researchers \texttt{(WHO)} focused on developing innovative ways of prototyping and making, these PDPs are built through a combination of predefined and open ended assembly, depending on the materiality their creators want to explore \texttt{(HOW)}. The innovative construction methods are able to 
support social engagement among the participants who bond over both shared and individual artefacts \texttt{(WHAT)}. The artefacts are built through creative tasks \texttt{(WHY)}, but structure the construction by following protocol based elicitation and by placing the PDPs in professional locations such as research labs or event venues \texttt{(WHEN/WHERE)}.


\subsubsection{\textbf{Validation}}

The primary agenda of \textcolor{valid}{\textit{Validation}} PDPs is to test and validate research hypotheses. These PDPs are designed to align with academic research visions \texttt{(WHERE/WHEN)} thus target scientists in their expert role in data visualisation \texttt{(WHO)}.
Participants are invited to engage with shared artefacts within research labs, where they perform both individual and collaborative data analysis \texttt{(WHAT)}. This analysis process is designed to facilitate the evaluation of tangible prototypes \texttt{(HOW)} through systematic tasks such as searching, filtering, and presenting \texttt{(WHY)}.

\section{Discussion}

Building on the overview of decisions presented across the 5 ontology dimensions (Section \ref{results section}), this Section discusses the most prominent power dynamics in PDP according to our definition, which defined PDP as a participatory method and data practice. 

From a participatory design perspective, stakeholders and participants who \textit{collaboratively contribute to PDP} 
expect to engage meaningfully, express themselves, and feel that their involvement can lead to a measurable impact. Without these foundations, the PDP process risks becoming extractive, soliciting input without reciprocating with meaningful benefits.
As participatory design scholars have consistently emphasised \cite{bodker_participatory_2018, bodker_tying_2017, bratteteig2012disentangling, huybrechts_institutioning_2017, teli_tales_2020, johnson_data-literacy_2021}, genuine participation prioritises transformative outcomes that promote resilience and self-determination by equipping participants with tools, knowledge, and social networks that extend beyond the scope of the participation.

From a data perspective, data is an artefact of power \cite{drucker2014graphesis} influenced by the decisions shaping how it is produced \cite{akbaba2024entanglements, loukissas_open_2021} and the agendas guiding its use \cite{dignazio_data_2020}. To meaningfully \textit{embody shared meanings in data}, participants need to have an understanding of the context where data is constructed and disseminated, and should be empowered to determine the extent of their involvement, as foregrounded by the contributions and labour of feminist scholars 
\cite{akbaba2024entanglements, meyer_criteria_2019, dignazio_data_2020, garcia_critical-refusal} and activists \cite{lupi2016dear,cifor2019feminist, costanza2018design}. 

We unpack these power dynamics by discussing how they influence the planning, deployment and construction process of PDP, according to the six PDP agendas outlined in Section \ref{agendas}, which highlight how varying orientations toward participation influence decision-making.

\subsection{Power dynamics through planning, deployment and dissemination of PDP} 

Our analysis reveals that even before participants begin constructing PDP, designers and other stakeholders make decisions that influence the construction and subsequent dissemination of a PDP artefact.
From a participatory design perspective, designers collaborate with stakeholders to address complex issues more effectively \cite{rossitto_efficiency_2021}. 
Including the power of representative organisations not only enhances the legitimacy of the process but also ensures broader ownership and support of collective outcomes \cite{huybrechts_institutioning_2017}. Such collaborations benefit stakeholders by allowing them to shape which issues are prioritised \cite{bodker_participation_2015} and how these issues are addressed \cite{frauenberger_pursuit_2015}. At the same time, designers draw on their expertise to guide the decision-making process \cite{bratteteig2012disentangling} and fulfil their responsibilities as project leaders \cite{bodker_participation_2015}. These power dynamics illustrate how power operates in PDP and inform the three key decisions that are typically taken before PDP construction begins.

\subsubsection{Power through planning decisions.}

According to our results (Section \ref{results-WHO}), when stakeholders and designers collaboratively plan PDP, the decisions are only shared among themselves, with participants generally having minimal involvement in the decision-making process.
A key example is selecting the appropriate elicitation activities and locations (Section \ref{results-WHEN}) for encoding meanings into data. These planning decisions are often informed by the need to match the data skills and interests of participants, ensuring that the process is meaningful and fosters a sense of agency \cite{kerzner_framework_2019}.

In participatory design processes, decisions about locations and participatory activities are often made without participant involvement due to practical constraints or pre-existing agendas. These decisions, described as originating from the `outside world' tend to be top-down and driven by local stakeholders who establish the framework within which participants are later invited to contribute into \cite{bratteteig2012disentangling}. 
However, since PDP is inherently a data practice, we argue that concealing these decisions from participants obscures the `hands that touched' the data \cite{dignazio_data_2020}: a lack of transparency regarding the settings where data was created can lead participants to provide inaccurate or misaligned contributions \cite{loukissas_open_2021}.

We believe that this power imbalance might have the greatest impact on PDPs with a \textcolor{pedagogy}{\textit{Pedagogy}} agenda, as their aim is exactly to improve data literacy among students (\textit{Data is Yours} \cite{bae_cultivating_2023}) and employees (\textit{SensorBricks} \cite{brombacher_sensorbricks_2024}), who develop skills in data collection, processing, and interpretation through their engagement in the PDP construction process. Critical data literacy research emphasises the importance of understanding the data context \cite{pangrazio2023critical} by highlighting that individuals only truly gain knowledge and skills when they first understand how data is produced and then apply their skills on issues relevant to them \cite{peer2019workshops, d2017creative}. To foster criticality among novice data users, researchers designed a learning programme involving participants in all decisions, including planning, allowing them to situate their skills and understanding in a real project \cite{disalvo_when_2024}, but the project stalled when they refused to continue until harmful practices they identified were addressed.

\textit{\textbf{PDP practices could involve participants more in planning decisions about locations and activities to enable them to contribute their issues more accurately.}} We believe that it is worthwhile to make planning decisions more transparent, aligning with the call for a more nuanced understanding of the subjective, narrative, and qualitative aspects of data \cite{van_koningsbruggen_what_2022}. This approach moves beyond interpreting data as purely objective and encourages
participants to more actively reflect on their understanding of data \cite{huron_lets_2017}. Naturally, such an approach may be challenging to implement due to various systemic constraints, such as the number of participants, organisational limitations, and logistical complexities.
However, a more practical approach to facilitating shared decision-making, inspired by visualisation research, involves clearly identifying the roles of those involved, as demonstrated by \cite{sedlmair2012design}.

\subsubsection{Power through decisions during deployment.}


According to our results (Sections \ref{results-WHAT} and \ref{results-WHO}), the deployment of PDPs often involves asking participants to embody issues using data that may lack relevance to their personal backgrounds.
For instance, PDPs with a \textcolor{valid}{\textit{Validation}} agenda are often deployed by designers to test research hypotheses, such as whether dynamic displays enhance information awareness in public spaces (\textit{CairnFORM} \cite{daniel_cairnform_2019}) or support collaborative data analysis (\textit{EMERGE} \cite{sturdee_exploring_2023}). Participants, who are often selected from immediate circles without a vested interest in the data being used, have little agency in choosing the data they are asked to encode. Instead, this data is typically predetermined and processed by the designers, as demonstrated in examples like \textit{Composites} \cite{subramonyam_composites_2022} and \textit{Wheeled Micro-Robots} \cite{le_goc_dynamic_2019}. 
This practice may arise from the limited skill set of participants in collecting or encoding data that address more complex issues, such as the availability of renewable energy (\textit{CairnFORM} \cite{daniel_cairnform_2019}).
While this power imbalance is less impactful for PDPs that place limited emphasis on participatory outcomes, including those with a \textcolor{valid}{\textit{Validation}} agenda, it poses greater challenges for PDPs with an \textcolor{action}{\textit{Action Research}} agenda where participants are only indirectly involved in selecting data. This could become particularly problematic when participants are expected to use the PDP artefacts that embody societal issues, such as water pollution (\textit{Chemicals in the Creek} \cite{perovich_chemicals_2021}), to organise and gain democratic influence over those issues.

\textit{\textbf{PDP practices could involve participants in decisions about their deployment, particularly in selecting data that aligns with their backgrounds and interests.}} We believe that making these decisions more transparent could enhance the outcomes of their participation, by embedding their real concerns \cite{iulia2020participation} about the issues they physically represent through the construction of PDP. Indeed, participants might still require assistance from designers to process data, particularly when defining data units or managing amplitude changes \cite{huron2014constructive}, which foregrounds why placing data selection entirely on participants is infeasible for all settings. 
However, building on the work of situated \cite{bressa_whats_2022} and personal \cite{huang2014personal} visualisation, which demonstrated the value of anchoring the creation of visualisations in community issues, this approach contributes to advancing data justice \cite{cifor2019feminist,costanza2018design}, an activist movement that advocates for data being used in ways that are ethical, equitable, and reducing harm among historically marginalised communities.

\subsubsection{Power through decisions about dissemination.}


Our results (Section \ref{results-WHY} and \ref{results-WHAT}) revealed that stakeholders and designers predetermine how PDP artefacts will be disseminated after their construction.
A key instance of a dissemination decisions is the selection of opportunistic locations and targeted communities where the PDP artefact can create a significant impact.
These decisions are perhaps guided by the goal of visualisations \cite{huang2014personal} to engage relevant audiences in their physical and social context, by bringing PDP artefacts closer to the highly-trafficked spaces \cite{valkanova2015public} and by constructing them through community-focused activities \cite{bressa_whats_2022}.

These decisions are best demonstrated by PDPs following \textcolor{practice}{\textit{Practice}} and \textcolor{engage}{\textit{Engagement}} agendas, as they are often meant to be used both during and after their construction.
These PDPs invite participants to assemble data
that embodies personal interests (see \textit{Cairn} \cite{gourlet_cairn_2017}, \textit{Craft Movement Translations} \cite{nissen_data-things_2015}) or personal experiences (see \textit{Small Talk} \cite{gallacher_smalltalk_2016} or \textit{KnowAndBe.Live} \cite{moretti_participatory_2020}), 
which are then meant to be shared with other participants as conference (\textit{Conference Tweet Translations} \cite{nissen_data-things_2015} and workshop badges (\textit{Data Badges} \cite{panagiotidou_data_2020}), as a public health awareness display (\textit{KnowAndBe.Live} \cite{moretti_participatory_2020}), or as a community practice dashboard (Cairn \cite{gourlet_cairn_2017}). 

Once issues are embodied in data represented by these physical artefacts, and are worn or communicated in authorable ways, these artefacts take on personal significance. As a result, participants become more invested in the dissemination process, feeling a sense of responsibility for the outcomes. However, this transfer of responsibility from stakeholders to participants can create tensions between stakeholder expectations and participant involvement \cite{frauenberger_pursuit_2015}.

\textit{\textbf{PDP practices could decide how to disseminate its artefacts in collaboration with participants.}} We consider it beneficial for participants to be informed from the outset about the intended use of their finalised artefact, as this would enable them to decide both the nature and extent of their participation.
Ideally, a transparently communicated assembly model could provide them with an agency to adapt the issues encoded in the artefact as they feel fit. This approach is illustrated by \textit{Data Badges} \cite{panagiotidou_data_2020}, where participants were allowed to reappropriate their artefacts by bypassing the encoding instructions to overcome privacy or sensitivity issues.

\subsection{Power dynamics during PDP construction}

Our analysis highlights that decisions about PDP construction are also shaped by designers and not entirely driven by participants.
As these decisions frame the design space in which the artefact can manifest within, 
designers must balance the flexibility and freedom of creative expression with the need to make the artefact easily understandable \cite{huron2014constructive}.
Furthermore, designers shape how participants collaboratively contribute to the artefact while establishing clear expectations for the outcome of their participation \cite{bodker_participatory_2018}, given that PDP artefacts are intended to embody shared meanings. These power dynamics inform two key critical decisions, which are examined further in this Section.

\subsubsection{Power through decisions about how to construct the PDP artefact.}

According to our results (see Sections \ref{results-HOW} and \ref{results-WHEN}), designers restrict the range of decisions participants can make to embody meanings into data during PDP construction.

Such decisions are reflected in the selection of hierarchical encoding channels, assembly models and elicitation activities that guide participants to embody their meanings into data.
These decisions are especially evident in \textcolor{engage}{\textit{Engagement}} PDPs, which are designed to produce artefacts that embody shared issues derived from personal data contributed by multiple collocated participants. In these cases, designers intentionally tend to simplify encoding decisions, reducing them to straightforward input tasks such as pressing buttons (\textit{SmallTalk} \cite{gallacher_smalltalk_2016}), rotating spinners (\textit{VoxBox} \cite{golsteijn_voxbox_2015}), or gliding plastic plates (\textit{Public visualisation Display} \cite{claes2015role}). This approach is often driven by the need for public-facing interfaces to support fast \cite{valkanova2014myposition} and intuitive \cite{valkanova2015public} interactions.

In contrast, PDPs following an \textcolor{explore}{\textit{Exploration}} agenda adopt a more open approach to encoding, by allowing participants greater flexibility in how they represent data. Participants are explicitly encouraged to explore unconventional strategies for embodying issues, such as via gamification (\textit{Co-gnito} \cite{panagiotidou_co-gnito_2022}), colourful and playful candy (\textit{Candy Workshop} \cite{lallemand_candy_2022}), or even knitting (\textit{Personal Data Narratives} \cite{friske_entangling_2020}).
However, despite their seemingly open-ended nature, these PDPs still predetermine the construction process through protocol-based elicitation activities. For example, participants are invited to enact predefined scenarios, such as simulating a business meeting by pulling elastic bands (\textit{Let's Get Physical} \cite{huron_lets_2017}), or place laser-cut tokens into a cafeteria dashboard following a prescribed data legend (\textit{Edo} \cite{sauve_edo_2023}).
We highlight the concern that these constraints limit the creative expression of personal issues. As such, participants may find it challenging to express personal food preferences (\textit{Edo} \cite{sauve_edo_2023}) or, even more so, to articulate emotions tied to personal experiences like walking on campus (\textit{Co-gnito} \cite{panagiotidou_co-gnito_2022}) or commuting from work (\textit{Personal Data Narratives} \cite{friske_entangling_2020}).

\textit{\textbf{PDP practices could provide more expressive freedom to support participants in constructing meaningful reflections with data.}}
We suggest that participants would benefit from expressing experiences in an open, qualitative way, acknowledging that participation involves not only rational decision-making but also subjective experiences \cite{frauenberger_scale_2018}. 
While trade-offs, such as the challenge of tracking excessive data categories, have been documented, emphasising self-expression through data holds potential for fostering reflection among participants \cite{thudt_self-reflection_2018}
By encouraging reflection, PDPs could overcome the limitations of empathy-driven approaches in participatory design, which have been criticised for reinforcing stereotypes and prioritising designers' views over participants' experiences \cite{bennett2019promise}.

\subsubsection{Power as consent during the PDP construction process.} \label{consent}


Our analysis (Section \ref{results-WHEN}, \ref{results-HOW}) revealed that participants often lacked the opportunity to dissent throughout the PDP construction process. As highlighted in Section \ref{sec:power in data}, feminist theory defines consent as a participant's right to make an informed decision, with the ability to change their mind at any time during participation. This includes being thoroughly informed about what they are consenting to \cite{garcia_critical-refusal} and having the freedom to withdraw their participation at any point without constraints \cite{zong2024data}, which were not made explicit during PDP construction.
For instance, participants may find it challenging to modify, recontextualise, or withdraw their contributions, even though they initially agreed to participate to the PDP construction. Best exemplified by PDPs with an \textcolor{engage}{\textit{Engagement}} agenda, participants contribute their personal data asynchronously, such as filling out surveys via quick interactions with tangible interfaces (\textit{VoxBox} \cite{golsteijn_voxbox_2015} and \textit{SmallTalk} \cite{gallacher_smalltalk_2016}), or by wrapping strings around pins on a board (\textit{KnowAndBe.Live} \cite{moretti_participatory_2020}). This asynchronous approach is common in public engagement visualisations, where the number of representative participants is too large to be recruited and co-located at a single location. Since the resulting PDP artefacts combine the distinct contributions of multiple participants, individual meanings are anonymised, making it difficult to trace them back to their original source.

One potential approach towards fostering consent is framed by \textcolor{practice}{\textit{Practice}} PDPs, where the underlying agenda allows participants to dissent at any point during the construction process.
Although participants embodied their meanings anonymously into individually constructed artefacts, such as \textit{Data Badges} \cite{panagiotidou_data_2020}, \textit{Conference Tweet Translations} \cite{nissen_data-things_2015}, or \textit{Craft Movement Translations} \cite{nissen_data-things_2015}, they were physically empowered to determine the level of private or sensitive data they wanted to encode.
In collectively constructed artefacts like \textit{Cairn} \cite{gourlet_cairn_2017} and \textit{VizBlocks} \cite{lechelt_vizblocks_2023}, participants were still able to visually identify their personal contributions because they were placed in dedicated physical spaces, which allowed them to remove or modify the data they encoded at any time.
While flexibility allows participants to alter or withdraw their physical representations, 
it can also lead to frictions and unintended outcomes as the audience may misinterpret the disseminated outcome. For example, in \textit{Data Badges} \cite{panagiotidou_data_2020}, several participants chose not to wear their data badges, or misconstrued them intentionally.

\textit{\textbf{PDP practices could design opportunities for expressing consent continuously throughout the construction process.}}
We argue that explicit consent throughout the construction process would encourage greater participation from members of vulnerable communities \cite{hodson2023whom, dai2021surfacing} and perhaps, make a step towards amplifying the voices that have been historically marginalised in conversations about data and technology \cite{hope2019hackathons, d2020personal, okerlund2021feminist}.
Alternatively, consent could be viewed as an evolving relationship between participants, stakeholders and designers, similar to what is proposed in the context of academic research \cite{garcia_critical-refusal}.

\section{Limitations}

All authors are working in the Western European context. Acknowledging her biases as a white woman with a background in participatory design, the first author realises that this analysis embraces feminist values such as social justice, transparency, accountability, empowerment, and care. 
The second author is a white woman with an academic research agenda focused on data visualisation, sustainability and AI.
The third author is a white woman with an academic research agenda in spatial analysis, mobility and spatial planning.
The last author is a white man whose academic agenda focuses on revealing the value of design in data visualisation, which foregrounds a Western appreciation of human creativity.

Regarding the development of the cross-disciplinary ontology, we acknowledge the subjectivity of the selection process of seed publications. While other papers may be better suited for this analysis, we believe that variations in theme definitions would not substantially alter the core meaning of the codes developed, and, as such, would not significantly affect the contributions of this study.

For the critical analysis of PDP, we selected PDP practices that were documented by academic literature publications. It is the rigour and transparency of their reporting that allowed us to analyse not only the PDP artefacts but also the decisions behind their design and construction. Therefore, we understand that it is perhaps unfair to build our critique on the best practices of PDP, considering that less rigorous practices remained obscure from our selection process. 

The intention of this analysis is not to undermine the PDP practice but to enhance it, as the work of this community already aligns closely with feminist values of transparency and empowerment. Furthermore, we believe that our recommendations, drawn from the exemplary practices of PDP, could provide a foundation for visualisation research in general to cultivate greater sensitivity toward fostering participation in collaborative knowledge production. 
However, we also acknowledge that our analysis does not fully embody the transparency we try to advocate, since we did not engage the authors of the analysed PDPs to meaningfully impact how their work was interpreted, nor did we offer them the opportunity to dissent to our claims.

As the issues we highlight stem from decisions shaped by contextually situated constraints and agendas of past PDP, we acknowledge that our recommendations are not sufficiently comprehensive to ensure entirely equitable PDP processes.
For instance, we realise that our recommendations cannot be applied to one PDP project in isolation, since they address issues that emerged across different agendas. Moreover, some recommendations may prove challenging to implement in the physical world, such as to allow participants to choose educational activities. 


\section{Conclusion}

Using a cross-disciplinary ontology that synthesises critical theoretical stances from participatory design and visualisation literature, we analysed a representative corpus comprising of 23 PDP artefacts, revealing a series of decisions and agendas in the PDP practice. 
By critically discussing the most prominent power dynamics among stakeholders and designers throughout the planning, deployment, and construction of PDPs, we argue that PDP should be considered a method that brings people together to collaboratively contribute to a physical representation that embodies shared issues in data, as PDP operates as both a participatory method and a data practice.
To better align PDP with feminist theory ideals of transparency, our study suggests
involving participants in the planning, deployment, and dissemination of PDPs to 
ensure that their issues are accurately represented; that the data 
reflects their backgrounds and interests; and that the artefact allows them to determine the nature and extent of their participation. Additionally, PDPs could provide participants with greater expressive freedom to construct meaningful reflections and to voice dissent throughout the construction process.
Collectively, our contributions form a foundational call for designers of data physicalisations and visualisations to reflect on and transparently share their decisions, making their processes of knowledge production more empowering for participants.

\begin{acks} \label{acks}

We thank the anonymous reviewers for their careful reading, insightful comments, and constructive suggestions, which have improved the clarity, rigour, and overall quality of this work.

We thank the authors of the PDP artefacts depicted in Figure \ref{fig:teaser} for giving us permission to use their images. The image credits are: 
1) \textit{Data is Yours }\cite{bae_cultivating_2023} img: S. Bae; 2)\textit{ CoDa} \cite{veldhuis_coda_2020} img: A. Veldhuis; 3) \textit{Sensor Bricks} \cite{brombacher_sensorbricks_2024} img: H. Brombacher; 4) \textit{Energy Blocks} \cite{panagiotidou_supporting_2023} img: G. Panagiotidou; 5) \textit{Chemicals in the Creek} \cite{perovich_chemicals_2021} img: W. Campbell; 6) \textit{VizBlocks} \cite{lechelt_vizblocks_2023} img: E. Morgan; 7) \textit{Craft Movement Translations} \cite{nissen_data-things_2015} img: B. Nissen; 8) \textit{Data Badges} \cite{panagiotidou_data_2020} img: G. Panagiotidou; 9) \textit{Cairn} \cite{gourlet_cairn_2017} img: P. Gourlet; 10) \textit{Conference Tweet Translations }\cite{nissen_data-things_2015} img: B. Nissen; 11) \textit{SmallTalk} \cite{gallacher_smalltalk_2016} img: Y. Rogers; 12) \textit{KnowAndBe.live} \cite{moretti_participatory_2020} \copyright 2020 Prevention For You S.r.L.; 13) \textit{VoxBox }\cite{golsteijn_voxbox_2015} img: Y. Rogers; 14) \textit{Public Visualisation Display} \cite{claes2015role} img: S. Claes;
15) \textit{Edo} \cite{sauve_edo_2023} img: K. Sauv\'e; 16) \textit{Co-gnito} \cite{panagiotidou_co-gnito_2022} img: G. Panagiotidou; 17) \textit{Personal Data Narratives} \cite{friske_entangling_2020} img: M. Friske; 18) \textit{Let's Get Physical} \cite{huron_lets_2017} CC BY 4.0 2017 S. Huron; 19) \textit{Candy Workshop} \cite{lallemand_candy_2022} img: C. Lallemand;
20) \textit{Cairnform }\cite{daniel_cairnform_2019} img: M. Daniel; 21) \textit{EMERGE} \cite{sturdee_exploring_2023} img: J. Alexander, 22) \textit{Composites} \cite{subramonyam_composites_2022} img: H. Subramonyam; 23) \textit{Wheeled Micro Robots} \cite{le_goc_dynamic_2019} img: M. Le Goc.

This  project  has  received  funding from the European Union’s Horizon 2020 research and innovation programme under the Marie Sklodowska-Curie grant agreement No. 955569. The opinions expressed in this document reflect only the authors' views and in no way reflect the European Commission’s opinions. The European Commission is not responsible for any use that may be made of the information it contains.
\end{acks}

\bibliographystyle{ACM-Reference-Format}
\bibliography{source/CHI25_Disentangling_TAPS}

\appendix

\section{Ontology}\label{sec:ontology}

This Section defines the ontology dimensions and describes the themes (e.g. \textbf{issue}), sub-themes (e.g. societal, personal) and codes (e.g. personal interests) of each dimension, as they appear listed in Table \ref{results}. 

\subsection{Who?} 

The \texttt{~WHO} dimension scrutinises the stakeholders who partake in the PD process. Tensions among stakeholders tend to arise due to unbalanced motivations, such as between personal interests, community values or institutional agendas \cite{bodker_participation_2015, bodker_rethinking_2016}. 
These stakeholders are typically distilled as established professional groups, with specific collaborator roles identified during the winnowing stage, based on their expertise in certain tasks and knowledge about possible misalignments \cite{sedlmair2012design}.

\begin{itemize}

    \item\textbf{participants} are bottom-up stakeholders who are actively involved in a PD process, driven by interests in issues of concern (e.g. residents, public members) \cite{frauenberger_pursuit_2015}, or users who perform tasks by following a research protocol (e.g. students, scientists) \cite{hornecker_design_2023,sauve_physecology_2022}, typically as part of a scientific study.
    
    \item\textbf{audience} refers to the individuals addressed by a physicalisation artefact according to their skills and knowledge, (e.g. general public, experts, children) \cite{bae_making_2022}.
    
    \item \textbf{stakeholders} have decision-making power in the PD process (e.g. non-profit, community, school). Their role in participation can change over the duration of a project \cite{frauenberger_pursuit_2015}.

    \item \textbf{designers} have the most power in the PD process, because they make or influence decisions at the project start, during implementation and throughout the PD process ~\cite{bratteteig2012disentangling}, that are shared with other stakeholders in different ways (e.g. designer in collaboration, designer by commission).
    
\end{itemize}

\subsection{What?}

The \texttt{~WHAT} dimension is sensitive to the inherent biases when collecting, processing and using data, while emphasising the risks of prioritising views of data as a neutral artefact \cite{dignazio_data_2020}.

\begin{itemize}
    \item \textbf{issue} addresses `matters of concern', which affect people in their everyday life \cite{latour2007reassembling}, and are evidenced with abstract data \cite{baibarac-duignan_controversing_2021} about societal issues \cite{dombrowski_social_2016} (e.g. environmental and health impact), and personal issues \cite{thorpe_design_2011} (e.g. personal interests).
    
    \item \textbf{data source} refers to the origin and type of the data, as objectively measured data (e.g. socio-economic) or more subjective (e.g. personal) \cite{bae_making_2022},
    while considering the methods and sources that add credibility to the dataset \cite{hullman2011visualisation}.

    \item \textbf{artefact} is the result of the PDP construction process \cite{huron_constructing_2014} (e.g. individual, shared).

    \item \textbf{data collection} involves the methods prompting participants to encode data \cite{huron_constructing_2014}, and they are either asynchronous (e.g. survey-based, sensor based) or synchronous (e.g. collaborative data analysis).
\end{itemize}

\subsection{Why?}

The \texttt{~WHY} dimension explores the social and political motivations \cite{bodker_tying_2017} that drive the decisions to use PDP as a PD method, knowing that it requires considerable effort from many stakeholders.

\begin{itemize}
    \item \textbf{output} involves the pragmatic and intangible artefacts produced by the PD process, as an early outcome \cite{hansen_how_2019, falk_what_2021}. Example of pragmatic artefacts are: action plan, used by participants to align on future actions \cite{hansen_how_2019}; learning, as the result of activities where participants develop new abilities \cite{hansen_how_2019}; prototype testing, as the process of developing a new technology iteratively, until it becomes a market-ready product \cite{hansen_how_2019}; evaluation results, as the findings obtained from assessing a technology \cite{hansen_how_2019}. Intangible artefacts are exemplified by problem analysis, where an ongoing process is iteratively improved to address a problem \cite{hansen_how_2019}; and social infrastructure, as ongoing relations developing among collaborating stakeholders \cite{hansen_how_2019}.
   
    \item \textbf{consequences} are short and mid-term outcomes from the PD process, including new work practices, new competences and/or skills \cite{hansen_how_2019, falk_what_2021} that give participants the ability to tackle issues.

    \item \textbf{impact} refers to the long-term outcomes of the PD process on participants, stakeholders and society at large, by improving their life quality, when the issues are related to everyday life \cite{hansen_how_2019}; by creating long-term relations, through alignment with other stakeholders and participants \cite{hansen_how_2019}; by creating social engagement, through social interactions with other stakeholders \cite{hansen_how_2019}; by gaining democratic influence, through personal engagement with societal issues \cite{hansen_how_2019}.
    
    \item \textbf{task} are performed by participants to encode meanings into data in both individual \cite{bae_making_2022} and collaborative contexts \cite{knoll_extending_2020}, which can be systematic (e.g. display and/or present; search, filter, analyse; input), or creative (e.g. plan, express).
\end{itemize}

\subsection{When and where?}

The \texttt{~WHEN/WHERE} dimension underpins the socio-political climate and historical and geographical positioning influencing participation \cite{huybrechts_visions_2020}, while having a critical understanding of the influence of community affiliations and organisational dynamics on participation \cite{sedlmair2012design}.

\begin{itemize}
    \item \textbf{community} includes the stakeholders with common agendas who self-organise based on shared goals (e.g. community of purpose \cite{clarke_socio-materiality_2021}) or based on having shared expertise in a specific domain (e.g. community of practice \cite{rossitto_efficiency_2021}, community of research). 
    
    \item \textbf{vision} refers to historical traditions, geographical conditions and political implications of the PD process \cite{huybrechts_visions_2020}, including knowledge visions (e.g. research, data literacy) and societal visions (economic, environmental, energy transition, maker culture).
    
    \item \textbf{location} includes a physical space, situated in physical boundaries and artefacts \cite{sauve_physecology_2022}; and the physical embodiment of the meanings \cite{bressa_whats_2022} attached to a place, while considering social and organisational values that are implicit in these places \cite{schoffelen2015visualising}. As such, PDP \textbf{locations} include professional locations (e.g. research lab, event venue) and opportunistic locations (e.g. office space, after-school, outdoor space).
    
    \item \textbf{elicitation activity} situates the PDP construction in relation to an activity that invites people to encode meanings into data \cite{kerzner_framework_2019}, which can happen spontaneously (e.g. physical interaction, digital interaction, play, ritual) or by following a protocol (e.g. demo training, scenario). 
    
\end{itemize}

\subsection{How?}

The \texttt{~HOW} dimension foregrounds the approaches, tools and techniques
used in the PD process \cite{sanders_co-creation_2008, sanders_probes_2014}, while considering the material affordances of the physicalisation artefact \cite{hogan_towards_2017}.

\begin{itemize}
    \item \textbf{design activity} involves the actions required to facilitate a PD process \cite{sanders_probes_2014}, including instructed activities (e.g. field study, evaluation) and informal activities (e.g. workshop, prototyping, infrastructuring i.e. the strategic process of creating and maintaining meaningful relations between stakeholders during the entire PD process \cite{bodker_tying_2017}).

    \item \textbf{participation methods} refer to the totality of approaches, tools and techniques chosen and combined by designers to involve participants in the PD process \cite{hansen_how_2019} including collaborative sense-making \cite{hansen_how_2019}, where participants are guided to develop new meanings together; mutual learning \cite{hansen_how_2019}, where participants learn from each other; and networking \cite{hansen_how_2019}, where stakeholders build relations through social interactions.

    \item \textbf{assembly model} points to the different activities and construction elements \cite{huron_constructing_2014} that are required to encode meanings into data, which can either be predefined, when the affordances, materials and assembly rules are determined before construction starts (e.g. tokens, blocks or strings; tangible prototype) or open-ended, where participants are co-determine these aspects during construction (e.g. material exploration).

    \item \textbf{encoding channel} refers to the physical variables that encode meanings into data \cite{hornecker_design_2023}, perceived in a hierarchical way (e.g. position and/or colour, volume, shape) or a non-hierarchical way (e.g. texture, motion and/or speed). 
       
\end{itemize}

\end{document}